\documentclass[conference]{IEEEtran}

\IEEEoverridecommandlockouts
\usepackage[T1]{fontenc}
\usepackage{microtype}
\usepackage{graphicx}
\usepackage{booktabs} % for professional tables
\usepackage{verbatim}
\usepackage{float}

\usepackage{cite}
\usepackage{hyperref}
\usepackage{algorithm}
\usepackage[noend]{algpseudocode}

\algdef{SE}[PROCEDURE]{Procedure}{EndProcedure}%
   [2]{\algorithmicprocedure\ #1\ifthenelse{\equal{#2}{}}{}{(#2)}}%
   {\algorithmicend\ \algorithmicprocedure}%

\usepackage{amsmath}
\usepackage{amssymb}
\usepackage{amsfonts}
\usepackage{mathtools}
\usepackage{amsthm}

\usepackage[capitalize,noabbrev]{cleveref}

\theoremstyle{plain}
\newtheorem{theorem}{Theorem}[section]
\newtheorem{proposition}[theorem]{Proposition}

\newtheorem{corollary}[theorem]{Corollary}
\theoremstyle{definition}

\theoremstyle{remark}

\usepackage[textsize=tiny]{todonotes}

\newcommand{\sysname}{CapCover Filter}

\AtBeginEnvironment{proof}{\parfillskip=0pt plus 1fil}
\AtBeginDocument{%
  \setlength{\abovedisplayskip}{6pt plus 2pt minus 2pt}%
  \setlength{\belowdisplayskip}{6pt plus 2pt minus 2pt}%
  \setlength{\abovedisplayshortskip}{2pt plus 2pt}%
  \setlength{\belowdisplayshortskip}{4pt plus 2pt minus 2pt}%
}
\begin{document}
\title{CapCover Filter: Semantic Approximate Membership Queries over Embeddings}

\author{\IEEEauthorblockN{Dvir David Biton\IEEEauthorrefmark{1},
Shay Hahiashvili\IEEEauthorrefmark{2},
Nadav Keren\IEEEauthorrefmark{2},
Gil Einziger\IEEEauthorrefmark{2}, and
Roy Friedman\IEEEauthorrefmark{1}}
\IEEEauthorblockA{\IEEEauthorrefmark{1}Technion, Haifa, Israel\\
Email: dbiton@campus.technion.ac.il, roy@cs.technion.ac.il}
\IEEEauthorblockA{\IEEEauthorrefmark{2}Ben Gurion University of the Negev, Be'er Sheva, Israel\\
Email: \{shayhah, nadavker\}@post.bgu.ac.il, gilein@bgu.ac.il}
}

\maketitle

\begin{abstract}
Determining whether an embedding is sufficiently similar to any vector in a vector store without performing a nearest-neighbor search can substantially improve query throughput in vector-store systems.
We study Semantic Approximate Membership Queries (SAMQs), which allow queries to be rejected when no indexed vector meets the requested similarity threshold, while guaranteeing no false negatives.

We introduce \sysname{}, a geometric data structure that summarizes a set of embeddings using spherical caps. Its representation is independent of the query threshold, allowing a single filter to support arbitrary thresholds without reconstruction. We evaluate alternative construction and quantization methods, compare \sysname{} with competing filters, and measure its impact on query throughput.

Across five real-world datasets, at a cosine similarity threshold of 0.95, \sysname{} eliminates approximately 20--100\% of index accesses while using only 2\% of the raw (FP32) embedding memory. When placed before a FAISS flat index, it improves query throughput by up to 382\% and by approximately 116\% on average across the five datasets, while returning the same range-search result. We also evaluate lower similarity thresholds. Our results show that \sysname{} can substantially improve vector-store throughput when similarity thresholds are tight, retrieval is expensive, and indexed embeddings are highly clustered. We release an optimized implementation of \sysname{} and its evaluation framework.
\end{abstract}


\section{Introduction}
\label{sec:intro}

Modern database and data-management systems increasingly use vector embeddings to represent and query unstructured data. Retrieval-Augmented Generation (RAG) retrieves passages from dense vector indexes~\cite{RAG-NLP}; vector database systems provide similarity search over high-dimensional representations~\cite{pan2024vectordbsurvey,Milvus}; and semantic caches compare incoming requests with previously served requests to determine whether a result can be reused~\cite{GPTCache,MeanCache,vCache,SemanticCaching}. Related uses of learned representations include semantic deduplication~\cite{SemDeDup} and large-scale recommendation systems~\cite{covington2016youtube}. Across these settings, semantic proximity often becomes part of a systems decision that is implemented through a vector search, even when the application needs only a binary indication of whether a sufficiently close vector exists.

\begin{figure}[htbp]
    \centering
    \includegraphics[width=0.8\columnwidth]{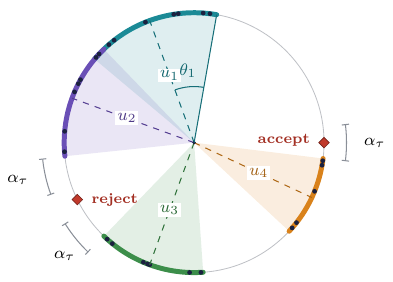}
    \caption{Embeddings on the unit sphere are covered by $k$ caps, each defined by a direction $u_i$ and an angle $\theta_i$. On query, each cap is conceptually widened by the angle $\alpha_\tau=\arccos(1-\tau)$ corresponding to the cosine-distance threshold $\tau$, resulting in an outer boundary of $\theta_i+\alpha_\tau$. A query vector is accepted for verification if it falls within at least one widened cap.}
    \label{fig:scheme}
\end{figure}

We focus on applications that need only this binary answer, such as a semantic cache that rejects a query as a miss once no
cached request meets its cosine similarity threshold, or a retrieval pipeline that screens queries before a costlier search or
verification stage. When negative queries are common, answering them cheaply avoids substantial downstream work.

For unit-normalized embeddings, let
$\operatorname{sim}_{\cos}(q,x)=\langle q,x\rangle$ denote cosine similarity.
Throughout the paper, $\tau\in[0,2]$ denotes the query's cosine-distance
threshold, corresponding to a minimum cosine similarity of $1-\tau$.
Given an indexed set $X\subseteq S^{d-1}$ and a unit query $q$, the exact
semantic-membership predicate is
\begin{equation}
\begin{aligned}
 I_X(q,\tau)
 &=\mathbf{1}\!\left[
     \min_{x\in X}\bigl(1-\operatorname{sim}_{\cos}(q,x)\bigr)\le\tau
   \right]
\end{aligned}
\label{eq:membership-pred}
\end{equation}

Classical Approximate Membership Query (AMQ) structures, such as Bloom filters and their relatives, summarize a set compactly and may
return false positives while preserving definite rejection of nonmembers~\cite{Bloom,carter1978membership,BloomFilterSurvey}.
The natural question is whether a similarly compact one-sided filter can support semantic proximity.

We refer to the semantic query analogue as a \emph{Semantic Approximate Membership Query (SAMQ)}.
A SAMQ returns \texttt{YES} for every query satisfying $I_X(q,\tau)=1$, while it may also return \texttt{YES} for some queries whose nearest indexed
vector has cosine similarity below $1-\tau$. 
Like a conventional one-sided AMQ, it trades false positives for a compact representation while preserving zero false negatives. 
This definition is independent of the representation used to implement~it.

Distance-sensitive membership is itself well established (\cref{sec:related-work}), but existing structures encode proximity
through locality-sensitive hashes and collisions, and similarity-search indexes retain point-level information needed to recover
neighbors. We instead store a bounded collection of enclosing regions that together contain all indexed vectors: a query
far from every stored region is rejected without examining any individual vector.

The geometric representation makes the coverage of the indexed set explicit, allowing us to impose a fixed physical memory budget directly on the filter. Each stored region may summarize many indexed vectors, and the number and precision of the regions can be chosen according to the available memory. As more vectors are summarized within the same budget, approximation appears as additional covered space, potentially increasing false positives, while the
covering invariant continues to preserve zero false negatives. Another useful property of the geometric representation is that its construction is independent of the query threshold. The stored regions depend on the indexed vectors and the representation budget, while $\tau$ appears only in the query-time acceptance condition. Consequently, the same constructed filter supports multiple cosine-distance thresholds.

More formally, embeddings are unit-normalized, so only their direction matters, and we realize SAMQs using a collection of at most $k$ spherical caps
\[
\begin{aligned}
    \mathcal{C}&=\{(u_1,\theta_1),\ldots,(u_k,\theta_k)\},\\
    C(u,\theta)&=\{x\in S^{d-1}:\angle(x,u)\le\theta\},
\end{aligned}
\]
whose union contains every indexed embedding, where $\angle(a,b)=\arccos\langle a,b\rangle$. We call this representation the \sysname{}. As illustrated in Figure~\ref{fig:scheme}, a query $(q,\tau)$ is accepted whenever
\[
    \angle(q,u_i)\le\theta_i+\alpha_\tau
\]
for some stored cap, where $\alpha_\tau=\arccos(1-\tau)$ is the angle corresponding to cosine distance $\tau$ on the unit sphere. Geometric containment and the triangle inequality for angles immediately imply that every query having cosine similarity at least $1-\tau$ to an indexed vector is accepted. False positives arise from the additional space admitted by the enclosing caps.

\begin{comment}
The effectiveness of the filter depends on how tightly the stored caps cover the indexed embeddings: excess covered space can admit queries whose cosine similarity to every indexed vector is below $1-\tau$ and therefore produce false positives. Writing $\sigma(\psi)$ for the normalized surface measure of a cap of angle $\psi$, the accepted region has measure at most
\[
J_\tau(\mathcal{C})=\sum_{i}\sigma(\theta_i+\alpha_\tau).
\]
This objective overcounts overlapping caps, but it is inexpensive to compute and upper-bounds the measure of the accepted region. We later relate it to the false-positive region and provide a distribution-dependent bound on its probability mass.
\end{comment}

Constructing a tight cover under a fixed memory budget is the central algorithmic problem. Our selected construction,
soft anneal, starts from Gonzalez's farthest-first directions~\cite{gonzalez1985clustering} followed by minimax refinement,
then attempts to reduce the total volume of the enclosing caps. We use reduced-precision directions, with angles re-measured so that
the covering invariant, and with it the zero-false-negative guarantee, is preserved.

\subsection*{Summary of Contributions}
\begin{itemize}
    \setlength{\itemsep}{2pt}
    \setlength{\parsep}{0pt}
    \item We present \sysname{}, a byte-budgeted geometric filter for SAMQs. Soft anneal constructs compact cap covers, while direction quantization and conservative angle remeasurement preserve coverage and zero false negatives (\cref{sec:samq,sec:static-construction,sec:correctness}).
    \item We prove coverage-based correctness, analyze query and construction costs, and relate false positives to the cover's geometric footprint and the query distribution. We also establish construction guarantees and a worst-case space lower bound for uniform queries (\cref{sec:construction-guarantees,sec:analysis}).
    \item Across five corpora, we evaluate construction and encoding choices, compare zero-false-negative filters at equal memory budgets, and measure throughput gains before an exact flat index. The results distinguish filtering selectivity from query cost, test the volume objective as an encoding-selection heuristic, and identify the backend-cost break-even condition (\cref{sec:evaluation}).
\end{itemize}

\section{Related Work}
\label{sec:related-work}

\subsection{Approximate and distance-sensitive membership}
Approximate membership structures represent a set compactly while allowing false positives, so that definite nonmembers
bypass more expensive operations~\cite{Bloom,carter1978membership}. Kirsch and Mitzenmacher's distance-sensitive Bloom filters
answer whether a query is close to some member using locality-sensitive hashing~\cite{kirsch2006distance}. The Locality-Sensitive
Bloom Filter (LSBF) studies the resulting false-positive and false-negative rates~\cite{hua2012lsbf}, and multi-granularity
filters support several query distances from one representation~\cite{qian2015mlbf,qian2016ilbf}. One-sided structures
have no false negatives: CoveringLSH guarantees a collision for every close pair in Hamming
space~\cite{pagh2016coveringlsh,goswami2017distance}, and no-false-negative locality-sensitive structures exist for $\ell_p$ metrics
and Euclidean near-neighbor search~\cite{ahle2017lasvegas,pacuk2016nofn,sankowski2017nofn}. 
Learned Bloom filters replace part of the
filter by a trained classifier and restore the one-sided guarantee with a backup filter~\cite{kraska2018learned,mitzenmacher2018learned}.
In contrast, \sysname{} needs no training and takes its guarantee from geometry. Range filters such as SuRF and Rosetta answer range emptiness for
keys~\cite{zhang2018surf,luo2020rosetta}, and filtered vector search restricts nearest-neighbor search by
attributes~\cite{gollapudi2023filtered}; neither decides whether any neighbor exists. These representations are hash- or point-oriented,
whereas the \sysname{} replaces the indexed observations at query time by a bounded conservative cover, and can be viewed as
approximate range emptiness~\cite{goswami2015range} for spherical caps.

\subsection{Similarity search and metric indexes}
ANN methods retain enough information to locate or approximate the distance to individual indexed vectors:
LSH~\cite{indyk1998ann,datar2004locality}, Multi-Probe LSH~\cite{lv2007multiprobe}, and graph-based indexes such as
HNSW~\cite{malkov2020hnsw}, whereas SAMQ may discard point identities after construction. Metric indexes (M-tree~\cite{ciaccia1997mtree}, cover trees~\cite{beygelzimer2006cover}) use regions to prune within a
point-level index.
Conversely, in a \sysname{} the regions are the whole query-time representation.

\subsection{Vector quantization and compressed representations}
Product Quantization (PQ)~\cite{jegou2011pq}, Optimized PQ~\cite{ge2013opq}, the scalar and PQ quantizers of
FAISS~\cite{FAISS}, and RaBitQ~\cite{gao2024rabitq} reduce the memory cost of high-dimensional similarity search.
These methods compress individual indexed vectors while retaining enough information for candidate
retrieval or distance estimation. Our \sysname{} applies quantization to the geometric representation itself:
a smaller direction code fits more caps into the same bytes and can reduce the angles of the cover, while the
displacement it introduces is accounted for conservatively so that coverage is never violated.

\subsection{Geometric covering, streaming summaries, and semantic caching}
The metric $k$-center problem controls the largest cluster radius, with Gonzalez's farthest-first algorithm being a factor-2
approximation under the triangle inequality~\cite{gonzalez1985clustering} (angular distance is such a metric). Other works
minimize the sum~\cite{gibson2010sumradii}, squares~\cite{friggstad2024radii}, or other norms~\cite{jaiswal2026capacitated} of the
radii. Our membership objective is related: in $d$ dimensions a cap's measure scales as $\sin^{d-1}\theta$, so the accepted
region is dominated by the widest caps, and the total measure of the caps is our tractable surrogate. BIRCH, streaming
clustering, and BICO keep bounded clustering summaries of a stream~\cite{zhang1996birch,ocallaghan2002streaming,fichtenberger2013bico}.
Semantic caches
(GPTCache~\cite{GPTCache}, MeanCache~\cite{MeanCache}, vCache~\cite{vCache}) decide whether reuse is valid and under what threshold.
SAMQ addresses the orthogonal question of whether definite misses can be identified from a compact representation first.

\section{\sysname{}}
\label{sec:samq}

We now describe the
geometric realization studied in this paper for the SAMQ abstraction introduced in Section~\ref{sec:intro}. 
Our goal is to represent
the region occupied by the indexed embeddings using a compact
collection of enclosing regions, while guaranteeing that every true
semantic match remains accepted.
Our implementation also supports insertions, but for the sake of brevity we restrict our presentation and evaluation to the static setting.

\subsection{Geometric Representation}
\label{sec:geometric-representation}

Write the indexed embeddings as $X=\{x_1,\ldots,x_n\}$. A \sysname{} stores a
collection $\mathcal C=\{C(u_i,\theta_i)\}_{i=1}^{m}$ of closed spherical caps,
where $m\le k$.

We say that $\mathcal C$ is a \emph{valid cover} of $X$ if
$X\subseteq\bigcup_{i}C(u_i,\theta_i)$, that is, for every indexed
embedding $x\in X$ there is a stored cap with $\angle(x,u_i)\le\theta_i$.

Recall the exact semantic-membership predicate $I_X(q,\tau)$ from
Section~\ref{sec:intro}. Given a query embedding $q$ and cosine-distance threshold
$\tau\in[0,2]$, the \sysname{} returns \texttt{YES} whenever some stored cap
satisfies $\angle(q,u_i)\le\theta_i+\alpha_\tau$, and returns \texttt{NO} otherwise. This query rule defines the behavior of our geometric realization. Other representations, such as a ball based filter, may implement the same one-sided
membership interface differently.

\begin{comment}    
The covering condition is sufficient to prevent false negatives. Consider an
indexed embedding $x$ with $\operatorname{sim}_{\cos}(q,x)\ge1-\tau$, that is,
$\angle(q,x)\le\alpha_\tau$. Because $\mathcal C$ covers $X$, some cap satisfies
$\angle(x,u_i)\le\theta_i$, and by the triangle inequality for angles
$\angle(q,u_i)\le\angle(q,x)+\angle(x,u_i)\le\alpha_\tau+\theta_i$, so the query
is accepted. Section~\ref{sec:correctness} formalizes this property.
\end{comment}

\subsection{Memory Budget and Geometric Objective}
\label{sec:cover-objective}

The physical memory cost of a \sysname{} depends on both the number of stored
caps and their representation. A cap requires a $d$-dimensional direction and a
single angle, making caps a parameter-efficient geometric primitive. The
numerical encoding of these parameters then determines how many caps can be
stored within a given memory budget. We write $M_{\mathrm{filter}}$ for
the filter's memory footprint and $M_{\mathrm{raw}}$ for the raw embedding
array's size, both in bytes, and often reference the ``compression ratio'' $M_{\mathrm{filter}}/M_{\mathrm{raw}}$.

Let $\mathcal B$ denote the total memory budget in bytes and let $E$ denote a
particular representation of the directions and angles. Let $s_E(d)$ denote the
number of bytes required per cap under encoding $E$, including its direction,
angle, and per-cap metadata, and let $h_E$ denote the number of bytes of
representation-wide state, such as shared quantization metadata or codebooks. The corresponding cap
budget is
\begin{equation}
\label{eq:cap-budget}
    k_E=\left\lfloor\frac{\mathcal B-h_E}{s_E(d)}\right\rfloor.
\end{equation}
Thus, $k$ is not itself a physical memory budget. Once the available memory and
representation are fixed, they induce the number of caps that can be stored. A
more compact representation may permit a larger $k$, while also introducing
numerical approximation into the stored directions and angles.

For a fixed representation, we seek a valid cover with small total cap volume. Let $\sigma(\psi)$ denote the normalized surface measure of a cap of angle $\psi$. We target the additive volume objective
\begin{equation}
\label{eq:cover-volume}
    V(\mathcal C)=\sum_i\sigma(\theta_i).
\end{equation}
This objective sums the volumes of the unexpanded caps containing the indexed embeddings, counting overlapping regions more than once. It is independent of the query threshold $\tau$. Although the union volume measures the covered region exactly, computing it is challenging, and bounds vary in tightness. We therefore use the additive objective as a tractable surrogate, which we empirically find to correlate with filtering effectiveness. We also use the maximum cap angle as a surrogate: in high dimensions, cap volume grows rapidly with angle below $\pi/2$, so the widest cap often dominates the additive objective.

We also use the expanded-cap measure
$J_\tau(\mathcal C)=\sum_i\sigma(\theta_i+\alpha_\tau)$, which upper-bounds the
measure of the accepted region. Section~\ref{sec:false-positive-analysis} relates
this quantity to false positives.

\begin{table}[t]
\centering
\caption{Main notation; specialized symbols are defined in the text.}
\label{tab:notation}
\footnotesize
\setlength{\tabcolsep}{4pt}
\begin{tabular}{@{}ll@{}}
\toprule
Symbol & Meaning \\
\midrule
$X$, $n$, $d$ & Indexed set, size, dimension \\
$q$, $X_i$ & Query vector; members assigned to cap $i$ \\
$\tau$ & Cosine-distance threshold \\
$\alpha_\tau$, $\delta_\tau$ & Angular and chord thresholds \\
$I_X(q,\tau)$ & Exact membership predicate \\
\midrule
$C(u_i,\theta_i)$ & Cap with direction $u_i$ and angle $\theta_i$ \\
$\mathcal C$, $m$, $k$ & Cover, cap count, cap limit \\
$R(\mathcal C)$, $R_k^\star(X)$ & Maximum cap angle; minimum with at most $k$ caps \\
$\widetilde u_i$, $e_i$ & Decoded direction; angular error \\
\midrule
$\mathcal B$, $E$, $k_E$ & Byte budget, encoding, cap budget \\
$s_E(d)$, $h_E$ & Bytes per cap; shared overhead \\
$M_{\mathrm{filter}}$, $M_{\mathrm{raw}}$ & Filter and raw-vector bytes \\
\midrule
$\sigma(\psi)$ & Normalized cap measure at angle $\psi$ \\
$V(\mathcal C)$, $J_\tau(\mathcal C)$ & Unexpanded and expanded volume sums \\
$\mu$, $\nu$ & Query law; uniform spherical measure \\
$U$, $P_\tau$, $A_\tau$ & Covered, true-positive, accepted regions \\
$\Delta_{\mathcal C}$ & Maximum distance to the nearest point in $X$ over $U$ \\
$F_\mu(t)$ & Query probability within angle $t$ of $X$ \\
\midrule
$N$, $\phi$ & Query count; fraction of accesses saved \\
$L_{\mathrm{filter}}$, $L$ & Average filter cost; constant backend cost \\
\bottomrule
\end{tabular}
\end{table}

\subsection{Static Cover Construction}
\label{sec:static-construction}

Our selected static construction uses soft anneal (\cref{sec:soft-anneal}). We
initialize the directions using Gonzalez's farthest-first algorithm~\cite{gonzalez1985clustering},
apply minimax refinement (\cref{sec:minimax-refinement}), and then use soft anneal
to improve the unexpanded cap-volume objective $V$. We then encode the resulting directions at
reduced precision and conservatively re-measure the stored angles to preserve
the covering condition. Note that following quantization, an embedding might end up closer to a different centroid than the one it was originally assigned to.

Because the encoding determines the cap budget in Equation~\ref{eq:cap-budget}, different encodings yield different covers under the same memory budget. Given a set $\mathcal E$ of candidate encodings, our default \texttt{auto} configuration constructs a cover $\mathcal C_E$ for each $E\in\mathcal E$ using its corresponding cap budget $k_E$ and a uniform encoding for all directions. After quantization and the volume-based reassignment check described below, it selects the representation with the smallest unexpanded cap-volume objective:
\[
E^\star=\arg\min_{E\in\mathcal E}V(\mathcal C_E).
\]
Section~\ref{sec:eval-quantization} evaluates this tradeoff. In the end-to-end and baseline experiments (\cref{sec:system,sec:eval-comparison}), the candidates are drawn from TQ (TurboQuant) and SPQ (Seeded Product Quantization) configurations, excluding other methods empirically dominated in compression and query throughput.

\subsubsection{Farthest-First Cover Construction}

For a given cap budget $1\le k\le n$, we initialize the cover using
Gonzalez's farthest-first algorithm under angular distance~\cite{gonzalez1985clustering}.
Starting from an arbitrary indexed embedding, the algorithm repeatedly selects
the embedding farthest from its nearest selected direction, until $k$
directions have been chosen. Each indexed embedding is then assigned to its
nearest direction. For the assigned set $X_i$, we set
$\theta_i=\max_{x\in X_i}\angle(x,u_i)$, yielding a valid cover by construction.

Because angular distance is a metric, farthest-first provides a factor-2
approximation to the minimum possible maximum cap angle
(\cref{thm:gonzalez-radius}). The traversal selects directions solely by distance,
without accounting for cluster density or the resulting total cap volume.
It can therefore allocate directions to isolated points while leaving
dense clusters covered by wide caps, motivating the refinements below.

\subsubsection{Minimax Refinement}
\label{sec:minimax-refinement}
Minimax refinement follows a Lloyd-style assignment-and-update
procedure~\cite{lloyd1982least}, replacing centroid updates with approximate
enclosing-cap fitting. Each round assigns every vector to its nearest
direction, then refines each cluster's direction using 64
B\u{a}doiu--Clarkson-style updates~\cite{badoiu2003coresets},
$c\leftarrow c+\frac{x_{\mathrm{far}}-c}{t+1}$, where $x_{\mathrm{far}}$
is the member farthest in angular distance from the current direction.
The resulting center is normalized to obtain a unit direction, and the
cap angle is remeasured over all assigned members.

We perform five rounds and retain the cover with the smallest maximum
cap angle, including the Gonzalez initialization as round zero.
Thus, before quantization, refinement cannot worsen the maximum-angle
objective relative to Gonzalez. We selected these constants empirically: they improved cover quality, while additional iterations
provided diminishing returns. Since construction is performed once
in our static setting, we prioritize cover quality over construction speed.

\subsubsection{Soft Anneal}
\label{sec:soft-anneal}
Soft anneal refines the minimax directions to reduce the unexpanded
cap-volume objective $V$. Writing $\angle_{xi}=\angle(x,u_i)$, we
softly assign each vector to its $\min\{16,k\}$ nearest directions using
$w_{xi}=\operatorname{softmax}_i\!\left(-\frac{\angle_{xi}}{T}\right)$,
refreshing the candidate lists every four iterations.
We optimize the smooth surrogate
\[
\begin{aligned}
 \Lambda_T&=\frac{1}{g}\log\sum_i\sigma(\hat\theta_i)^g,\\
 \hat\theta_i&=T\log\frac{\sum_x w_{xi}e^{\frac{\angle_{xi}}{T}}}
                            {\sum_x w_{xi}}.
\end{aligned}
\]
The softmax is restricted to each vector's candidate list, and the
sums over $x$ include only vectors listing direction $i$.
We do not explicitly weight cap contributions by their member counts.
Although favoring tighter caps in denser regions improved FPR in our
experiments, this benefit may reflect the similarity between the query
and indexed-data distributions. We omit such weighting to avoid assuming
that indexed-data density predicts query density in other workloads.

A Riemannian Adam variant~\cite{becigneul2019riemannian}, with
$(\beta_1,\beta_2)=(0.9,0.99)$, updates the unit directions using
tangent gradients, normalization retraction, and projection transport
of momentum. Over 300 iterations, $T$, the learning rate, and $g$
follow geometric schedules from $0.1$ to $0.002$, $0.1$ to $0.005$,
and $\frac{30}{d-1}$ to $1$, respectively.
Every ten iterations, we evaluate $V$ using hard nearest-direction
assignments and maximum member angles.

We retain the best evaluated iterate, including the minimax
initialization, and accept a final minimax round with 64
B\u{a}doiu--Clarkson-style updates only if it further reduces $V$.
Before quantization, the returned cover therefore has no larger $V$
than the minimax initialization. Measuring cap angles over all assigned
members preserves coverage (\cref{thm:no-fn}).

Soft anneal takes substantially longer than the preceding stages.
It is appropriate when additional construction time is available to
improve the volume objective; minimax alone offers a less expensive
alternative.

\subsubsection{Quantization}
\label{sec:quantized-cover}

After constructing the cover, we encode all cap directions using a single
reduced-precision encoding $E$. Let
$\theta_i^0=\max_{x\in X_i}\angle(x,u_i)$ denote the angle of cap $i$
before encoding, and let $\widetilde u_i$ be its decoded unit direction.
Keeping the assigned members $X_i$ fixed, we remeasure the cap angle as
\[
 \widetilde\theta_i^{\rm fix}
 =\max_{x\in X_i}\angle(x,\widetilde u_i)
 \le\theta_i^0+e_i,
\]
where $e_i=\angle(u_i,\widetilde u_i)$ is the angular encoding error.
The resulting cover $\mathcal C_E^{\rm fix}$ preserves coverage by
Proposition~\ref{prop:quantization-coverage}.

We then tentatively reassign every indexed vector to its nearest decoded
direction, remeasure the cap angles over their new members, and discard
empty caps. This produces another valid cover, $\mathcal C_E^{\rm re}$.
We retain the reassignment only if it strictly reduces the unexpanded
cap-volume objective:
\[
 V(\mathcal C_E^{\rm re})<V(\mathcal C_E^{\rm fix}).
\]
Otherwise, we retain the original assignments with their remeasured angles.
Both candidates are evaluated using the angles that would be stored,
rounded outward as needed to preserve coverage. Thus, either choice
preserves coverage, and the retained cover has no larger $V$ than
$\mathcal C_E^{\rm fix}$.

We do not compress cap angles since each requires only one scalar, so we
focus compression on the $d$-dimensional directions. Each angle is measured
around the decoded direction, so it absorbs the direction's quantization error,
and it is rounded upward when stored so that coverage is preserved. We also evaluated
selecting an encoding separately for each cap, but the small FPR
improvement did not justify the loss in query throughput.

\begin{algorithm}[t]
\caption{\sysname{} Construction}
\label{alg:static}
\small
\begin{algorithmic}[1]
\Require Nonempty indexed set $X$, byte budget $\mathcal B$,
         candidate encodings $\mathcal E$ with at least one feasible encoding
\State $\mathcal E_{\rm feasible}\gets\emptyset$
\For{each encoding $E\in\mathcal E$}
    \State Compute $k_E$ from Equation~\ref{eq:cap-budget}
    \If{$k_E\ge1$}
        \State Add $E$ to $\mathcal E_{\rm feasible}$
        \State Initialize $\min\{k_E,|X|\}$ directions by farthest-first
        \State Apply minimax refinement (\cref{sec:minimax-refinement})
        \State Apply soft anneal to improve $V$ (\cref{sec:soft-anneal})
        \State Assign each $x\in X$ to its nearest direction, forming $X_i$
        \For{each direction $u_i$ with nonempty $X_i$}
            \State Encode $u_i$ using $E$ and decode its unit direction $\widetilde u_i$
            \State $\widetilde\theta_i^{\rm fix}\gets
                \max_{x\in X_i}\angle(x,\widetilde u_i)$
            \State Round $\widetilde\theta_i^{\rm fix}$ outward as needed
        \EndFor
        \State Let $\mathcal C_E^{\rm fix}$ be the resulting valid cover
        \State Reassign each $x\in X$ to its nearest decoded direction
        \State Remeasure cap angles over the new members
        \State Round the remeasured angles outward as needed
        \State Let $\mathcal C_E^{\rm re}$ be the resulting valid cover
        \State $\mathcal C_E\gets\mathcal C_E^{\rm fix}$
        \If{$V(\mathcal C_E^{\rm re})<V(\mathcal C_E^{\rm fix})$}
            \State $\mathcal C_E\gets\mathcal C_E^{\rm re}$
        \EndIf
    \EndIf
\EndFor
\State $E^\star\gets
    \arg\min_{E\in\mathcal E_{\rm feasible}}V(\mathcal C_E)$
\State \Return $\mathcal C_{E^\star}$
\end{algorithmic}
\end{algorithm}

For $m$ stored caps, a query scan costs $O(md)$, plus any
encoding-specific preprocessing, such as rotation or lookup-table
construction. A dense rotation adds $O(d^2)$ work per query.
Farthest-first construction and each full nearest-direction
assignment pass cost $O(nmd)$; total construction cost depends
on the refinement iterations and candidate encodings.

Algorithm~\ref{alg:static} summarizes the static construction.

\section{Theoretical Analysis}
\label{sec:analysis}

\subsection{Setting and notation}
We assume a nonempty indexed set $X\subseteq S^{d-1}$ with $d\ge2$,
unit query vectors, and cosine-distance threshold $\tau\in[0,2]$.
Angular distance is a metric. With $\alpha_\tau=\arccos(1-\tau)$,
\begin{equation}
\label{eq:alpha}
\begin{aligned}
 \angle(q,x)\le\alpha_\tau
 &\iff\operatorname{sim}_{\cos}(q,x)\ge1-\tau\\
 &\iff 1-\operatorname{sim}_{\cos}(q,x)\le\tau.
\end{aligned}
\end{equation}
The equivalent Euclidean chord threshold is
$\delta_\tau=\sqrt{2\tau}$, giving
$\alpha_\tau=2\arcsin\!\left(\frac{\delta_\tau}{2}\right)$.
We use angular distance unless Euclidean distance is explicitly indicated.

We interpret cap angles at least $\pi$ as covering the entire sphere.
Throughout this section, $\mathcal C$ consists of decoded unit directions
$\widetilde u_i$ and stored angles $\theta_i\in[0,\pi]$.
Writing $U=\bigcup_i C(\widetilde u_i,\theta_i)$, validity means
$X\subseteq U$. The filter accepts a query exactly when
\[
 \angle(q,\widetilde u_i)
 \le\min\{\pi,\theta_i+\alpha_\tau\}
 \quad\text{for some }i.
\]
The analysis assumes exact arithmetic and this acceptance rule.

For a query distribution $\mu$ and filter $\Gamma$, define
\[
\begin{aligned}
 &\mathrm{FPR}_\mu(\Gamma,\tau)\\
 &\quad=\Pr_{q\sim\mu}\!\left[
 \Gamma(q,\tau)=\texttt{YES}\mid I_X(q,\tau)=0\right],
\end{aligned}
\]
provided that negative queries have positive probability.
We also use a cover to denote the filter it induces under the
specified decoding rule.

Let $\nu$ denote the uniform probability measure on $S^{d-1}$.
The normalized cap measure $\sigma(\psi)=\nu(C(u,\psi))$ is
independent of $u$ and satisfies
\[
 \sigma(\psi)=
 \frac{\int_0^\psi\sin^{d-2}s\,ds}
      {\int_0^\pi\sin^{d-2}s\,ds}
 \quad(0\le\psi\le\pi).
\]
We extend $\sigma$ as zero for $\psi\le0$ and one for $\psi\ge\pi$,
and write $R(\mathcal C)=\max_i\theta_i$ for the maximum cap angle.

The following bounds quantify how cap angles affect covered surface
area. For $0<\psi<\frac{\pi}{2}$, the substitution
$\sin s=t\sin\psi$ shows that
$\frac{\sigma(\psi)}{\sin^{d-1}\psi}$ is nondecreasing.
By continuity and $\sigma\!\left(\frac{\pi}{2}\right)=\frac12$,
\[
 \sigma(\psi)\le\frac12\sin^{d-1}\psi
 \quad\left(0\le\psi\le\frac{\pi}{2}\right).
\]
For $0<a\le b<\frac{\pi}{2}$, the same substitution yields
\begin{equation}
\label{eq:ratio}
 \left(\frac{\sin b}{\sin a}\right)^{d-1}
 \le\frac{\sigma(b)}{\sigma(a)}
 \le\frac{1}{\cos b}
       \left(\frac{\sin b}{\sin a}\right)^{d-1}.
\end{equation}
The lower bound follows from the preceding monotonicity.
The upper bound follows by bounding the transformed integrand's
factor $(1-t^2\sin^2\psi)^{-1/2}$ between $1$ and
$\frac{1}{\cos\psi}$.

\subsection{Correctness}
\label{sec:correctness}

\begin{theorem}[No false negatives]
\label{thm:no-fn}
Every valid cover accepts every query with $I_X(q,\tau)=1$.
\end{theorem}
\begin{proof}
Choose $x\in X$ with $\angle(q,x)\le\alpha_\tau$ and a cap
containing $x$. The angular triangle inequality gives
$\angle(q,\widetilde u_i)\le
\angle(q,x)+\angle(x,\widetilde u_i)\le\alpha_\tau+\theta_i$.
\end{proof}

\begin{proposition}[Exact cap neighborhood]
\label{prop:exact-neighbourhood}
A query is within cosine distance $\tau$ of some point in
$C(u,\theta)$ if and only if
$\angle(q,u)\le\min\{\pi,\theta+\alpha_\tau\}$.
\end{proposition}
\begin{proof}
The minimum angular distance from $q$ to the cap is
$\max\{0,\angle(q,u)-\theta\}$; apply~\eqref{eq:alpha}.
\end{proof}

Thus the accepted region is exactly the angular
$\alpha_\tau$-neighborhood of $U$.

\begin{proposition}[Quantization preserves coverage]
\label{prop:quantization-coverage}
For a nonempty assigned set $X_i$, let
$\widetilde\theta_i=\max_{x\in X_i}\angle(x,\widetilde u_i)$.
With $\theta_i^0$ and $e_i$ as in \cref{sec:quantized-cover},
$\widetilde\theta_i\le\theta_i^0+e_i$.
Any stored angle $\theta_i\ge\widetilde\theta_i$ covers $X_i$,
including an angle encoded on a grid and rounded outward.
\end{proposition}
\begin{proof}
Apply the angular triangle inequality to each member of $X_i$.
\end{proof}

\subsection{Comparison with Ball Covers}
\label{sec:ball-comparison}
Let $\mathcal D=\{B_2(c_i,r_i)\}_{i=1}^m$ cover $X$.
Euclidean expansion accepts
\[
 A_{\rm ball}(\tau)=\bigcup_i
 \bigl(B_2(c_i,r_i+\delta_\tau)\cap S^{d-1}\bigr).
\]
Each nonempty intersection $B_2(c_i,r_i)\cap S^{d-1}$ is a spherical
cap, possibly the entire sphere. These intersections define a valid
cover $\mathcal C(\mathcal D)$ with at most $m$ caps.
Let $A_{\rm cap}(\tau)$ denote the region accepted by
$\mathcal C(\mathcal D)$ under angular decoding.

\begin{theorem}[Angular decoding of a ball cover]
\label{thm:cap-vs-ball}
For every ball cover $\mathcal D$ and $\tau\in[0,2]$,
\[
 A_{\rm cap}(\tau)\subseteq A_{\rm ball}(\tau).
\]
Consequently, whenever FPR is defined,
\[
 \mathrm{FPR}_\mu(\mathcal C(\mathcal D),\tau)
 \le \mathrm{FPR}_\mu(\mathcal D,\tau).
\]
\end{theorem}
\begin{proof}
An accepted cap query $q$ is within chord distance $\delta_\tau$
of some $y\in B_2(c_i,r_i)\cap S^{d-1}$. Thus
$\|q-c_i\|_2\le\|q-y\|_2+\|y-c_i\|_2\le\delta_\tau+r_i$.
\end{proof}

The excess accepted area can be substantial even for a ball
whose spherical intersection is exactly the cap being expanded.
For $0<\theta<\frac{\pi}{2}$, the ball
$B_2(\cos\theta\,u,\sin\theta)$ intersects the sphere exactly in
$C(u,\theta)$.

\begin{corollary}[Expansion overhead]
\label{cor:ball-expansion-overhead}
Let $0<\theta<\frac{\pi}{2}$, $0<\alpha=\alpha_\tau$, and
$\gamma=\theta+\alpha<\frac{\pi}{2}$.
Euclidean expansion of this ball accepts $C(u,\beta)$, where
\[
 \beta=\arccos\!\left(\max\left\{-1,
 \cos\theta-\delta_\tau\tan\theta-\tau\sec\theta
 \right\}\right)>\gamma.
\]
Its accepted surface measure satisfies
\[
 \frac{\sigma(\beta)}{\sigma(\gamma)}
 \ge
 \left(
 \frac{\sin(\min\{\beta,\frac{\pi}{2}\})}{\sin\gamma}
 \right)^{d-1},
\]
which grows exponentially with $d$ for fixed $\theta$ and $\alpha$.
\end{corollary}
\begin{proof}
Expanding
$\|q-\cos\theta\,u\|_2^2\le(\sin\theta+\delta_\tau)^2$
gives $\beta$. Before clipping, its cosine threshold is smaller
than $\cos\gamma$ by
\[
 \delta_\tau\tan\theta
 \left[1-\cos\theta\cos\!\left(\frac{\alpha}{2}\right)\right]
 +\tau(\sec\theta-\cos\theta)>0.
\]
The measure bound follows from~\eqref{eq:ratio} and monotonicity.
\end{proof}

This comparison concerns angular versus Euclidean expansion of
the same ball--sphere intersections. A ball decoder using their
exact spherical neighborhoods matches the converted cap decoder.
The single-cap surface-measure ratio does not imply the same ratio
for unions of caps or workload FPR.

\subsection{Construction Guarantees}
\label{sec:construction-guarantees}

Let $R_k^\star(X)$ denote the minimum maximum angle of a valid
cover with at most $k$ caps.

\begin{theorem}[Farthest-first angular guarantee]
\label{thm:gonzalez-radius}
For $1\le k\le n$, farthest-first with nearest-direction assignment
produces a cover $\mathcal C_G$ satisfying
\[
\begin{aligned}
 R(\mathcal C_G)&\le2R_k^\star(X),\\
 J_\tau(\mathcal C_G)&\le
 k\sigma(2R_k^\star(X)+\alpha_\tau).
\end{aligned}
\]
\end{theorem}
\begin{proof}
If the final radius is $R>0$, the $k$ selected points and a point
attaining $R$ are pairwise separated by at least $R$.
Two belong to the same cap of an optimal cover, so
$R\le2R_k^\star(X)$. Monotonicity of $\sigma$ gives the second
bound. Both bounds also hold for $R=0$.
\end{proof}

Farthest-first can incur a volume penalty exponential in dimension.
For $k\ge2$, place $k$ pairs at $\frac{2\pi j}{k}\pm a$ on a great
circle, where $j=0,\ldots,k-1$ and $0<a<\frac{\pi}{2k}$.
Since inter-pair distances exceed $2a$, farthest-first selects one
endpoint per pair regardless of initialization or tie-breaking.
Nearest-direction assignment then produces $k$ caps of angle $2a$.
Caps centered at the pair midpoints instead cover the points with
angle $a$. Thus, if $V_k^\star(X)$ denotes the minimum additive
volume of a valid cover with at most $k$ caps,
\[
 \frac{V(\mathcal C_G)}{V_k^\star(X)}
 \ge\frac{\sigma(2a)}{\sigma(a)}
 \ge(2\cos a)^{d-1}.
\]
Hence the factor-two angle guarantee does not imply a
dimension-independent volume approximation.

\subsection{Geometric Footprint and False Positives}
\label{sec:false-positive-analysis}

Let
\[
 P_\tau=\bigcup_{x\in X}C(x,\alpha_\tau),
 \qquad
 A_\tau=\bigcup_iC(\widetilde u_i,\theta_i+\alpha_\tau)
\]
denote the true-positive and accepted regions. Validity gives
$P_\tau\subseteq A_\tau$.

\begin{proposition}[Accepted mass and false positives]
\label{thm:false-positive-mass-bound}
For any query distribution $\mu$ with $\mu(P_\tau)<1$,
\[
 \mathrm{FPR}_\mu(\mathcal C,\tau)
 =\frac{\mu(A_\tau)-\mu(P_\tau)}{1-\mu(P_\tau)}
 \le\mu(A_\tau).
\]
In particular,
$\mathrm{FPR}_\nu(\mathcal C,\tau)\le\min\{1,J_\tau(\mathcal C)\}$.
More generally, if $\mu$ is absolutely continuous relative to $\nu$
with density at most $H$ almost everywhere, then
$\mathrm{FPR}_\mu(\mathcal C,\tau)\le\min\{1,HJ_\tau(\mathcal C)\}$.
\end{proposition}
\begin{proof}
For $a=\mu(A_\tau)$ and $p=\mu(P_\tau)$,
$a-p\le a(1-p)$ since $a\le1$.
The remaining bounds follow from the union bound and the density bound.
\end{proof}

\begin{corollary}[Clustered data]
\label{cor:clustered}
\label{thm:clustered}
If $X$ lies in at most $k$ caps of angle $\rho$, with $1\le k\le n$,
then its unquantized farthest-first cover $\mathcal C_G$ satisfies,
whenever FPR is defined and $2\rho+\alpha_\tau\le\frac{\pi}{2}$,
\[
 \mathrm{FPR}_\nu(\mathcal C_G,\tau)
 \le\min\left\{1,\frac{k}{2}
 \sin^{d-1}(2\rho+\alpha_\tau)\right\}.
\]
\end{corollary}
\begin{proof}
The assumed cover gives $R_k^\star(X)\le\rho$, hence
$R(\mathcal C_G)\le2\rho$. Apply the preceding proposition
and the cap-measure bound.
\end{proof}

These bounds describe the unquantized farthest-first cover, not the soft-anneal covers evaluated in \cref{sec:evaluation}. This bound decays exponentially with $d$ when
$2\rho+\alpha_\tau\le\psi<\frac{\pi}{2}$ for fixed $\psi$
and $\log k=o(d)$. These bounds concern expanded caps;
reducing the unexpanded objective $V$ does not by itself
guarantee lower workload FPR. For arbitrary query distributions, we can instead bound false
positives by the probability mass near the true matching boundary.

\begin{theorem}[Cover-excess shell bound]
\label{thm:cover-shell}
For the stored region $U$, define
\[
\begin{aligned}
 \Delta_{\mathcal C}
 &=\sup_{u\in U}\min_{x\in X}\angle(u,x),\\
 F_\mu(t)
 &=\Pr_{q\sim\mu}\!\left[\min_{x\in X}\angle(q,x)\le t\right].
\end{aligned}
\]
Whenever $F_\mu(\alpha_\tau)<1$,
\[
 \mathrm{FPR}_\mu(\mathcal C,\tau)
 \le
 \frac{
 F_\mu(\min\{\pi,\alpha_\tau+\Delta_{\mathcal C}\})
 -F_\mu(\alpha_\tau)}
 {1-F_\mu(\alpha_\tau)}.
\]
\end{theorem}
\begin{proof}
Every accepted query is within angle $\alpha_\tau$ of some
$u\in U$, which is within $\Delta_{\mathcal C}$ of an indexed point.
Thus every false positive lies in the shell
\[
 \alpha_\tau<\min_{x\in X}\angle(q,x)
 \le\min\{\pi,\alpha_\tau+\Delta_{\mathcal C}\}.
\]
Conditioning its probability on a negative query proves the bound.
\end{proof}

The quantity $\Delta_{\mathcal C}$ measures the cover's excess
angular reach and is independent of the query threshold.
If each cap contains its nonempty assigned set $X_i$, the triangle
inequality gives
\[
\begin{aligned}
 \Delta_{\mathcal C}
 &\le\min\left\{\pi,\max_i\left[
 \theta_i+\min_{x\in X_i}\angle(\widetilde u_i,x)\right]\right\}\\
 &\le\min\{\pi,2R(\mathcal C)\}.
\end{aligned}
\]
Thus the same cover-excess bound applies across all thresholds,
although it can be loose when the cover contains wide caps.

\subsection{Space Lower Bound}
\label{sec:space-lower-bound}

We derive a representation-independent bound for deterministic filters
over the continuous sphere. All persistent data-dependent information
occupies a fixed array of $\mathcal B$ bytes; the decoder accesses only
this state, the query, and fixed parameters. Encoding and query rules
are measurable. The proof combines spherical isoperimetry with
finite-state counting, following ideas from similarity
identification~\cite{ingber2015quadratic,steiner2016quadratic}.

\begin{theorem}[Uniform-sphere space lower bound]
\label{thm:info-lower-bound}
Fix $d\ge2$, $n\ge1$, $0<\alpha=\alpha_\tau<\pi$, and
$0<\varepsilon<1$, with
\[
 v=\sigma(\alpha),\qquad
 \eta=\varepsilon+(1-\varepsilon)nv<1.
\]
Every deterministic structure representing all $n$-point subsets
of $S^{d-1}$, with no false negatives and conditional uniform-query
FPR at most $\varepsilon$ for every dataset, requires
\[
 \boxed{\displaystyle
 \mathcal B\ge\frac{n}{8}\log_2
 \frac{1}{\sigma(\sigma^{-1}(\eta)-\alpha)}.}
\]
Here $\sigma^{-1}$ is the inverse cap-measure function on $[0,\pi]$.
\end{theorem}

\begin{proof}
For each used state $\omega$, let $A_\omega$ be its accepted region.
For any dataset represented by this state, the FPR guarantee gives
\[
 \nu(A_\omega)
 \le\varepsilon+(1-\varepsilon)\nu(P_\tau)
 \le\eta.
\]
No false negatives imply that every represented point belongs to
\[
 G_\omega=
 \{x:\operatorname{dist}_{\angle}(x,A_\omega^c)\ge\alpha\}.
\]
For $0<\xi<\alpha$, the angular $\xi$-neighborhood of $G_\omega$
lies in $A_\omega$. Spherical isoperimetry
\cite[Lemma~2, journal version]{ingber2015quadratic},
followed by $\xi\uparrow\alpha$, therefore yields
\[
 \nu(G_\omega)\le
 p:=\sigma(\sigma^{-1}(\eta)-\alpha).
\]
An ordered tuple of $n$ independent uniform points has distinct
entries almost surely and belongs to $G_\omega^n$ for some state.
Each such product set has measure at most $p^n$, and there are at most
$2^{8\mathcal B}$ states. Hence $1\le2^{8\mathcal B}p^n$.
\end{proof}

When $nv\le\varepsilon\le1/8$, we have $\eta\le2\varepsilon$.
Writing $\beta=\sigma^{-1}(\eta)$, we have
$\alpha_\tau<\beta<\frac{\pi}{2}$, and the cap-measure bound gives
\[
\begin{aligned}
 p&=\sigma(\beta-\alpha_\tau)\\
 &\le\min\left\{2\varepsilon,\frac12\cos^{d-1}\alpha_\tau\right\}\\
 &\le\min\left\{2\varepsilon,
 \frac12 e^{-\frac{(d-1)\alpha_\tau^2}{2}}\right\}.
\end{aligned}
\]
Consequently,
\[
 \mathcal B=\Omega\!\left(
 n\left[\log\frac{1}{\varepsilon}+(d-1)\alpha_\tau^2\right]
 \right)\quad\text{bytes}.
\]
Here $\alpha_\tau^2=\Theta(\tau)$ as $\tau\downarrow0$.
As an example, for $d=384$, $\tau=0.02$, $n=80{,}000$, and
$\varepsilon=0.01$, the theorem gives approximately $4.08$ bytes
per indexed vector.

This bound requires the FPR guarantee for every dataset under
uniform queries. It does not bound the space needed for a particular
clustered dataset or an empirical workload.

\section{Evaluation}
\label{sec:evaluation}
Our evaluation examines how much work a compact filter can save before a query reaches the vector index. We build filters under different memory budgets and test them on queries whose correct answers are established by exact search. We measure both how often the filter avoids an index lookup and whether the saved work outweighs the cost of filtering. Additional plots, raw data, and the evaluation framework are available online~\cite{GitHub}.

\subsection{Experimental Setup}
\label{sec:eval-setup}

\subsubsection{Workloads}
We evaluate \sysname{} on five datasets that feature paraphrased or semantically similar data:
\textbf{WildChat}~\cite{WildChat}, user prompts to LLM chat services;
\textbf{CC-News}~\cite{CCNews}, lightly rewritten news articles;
\textbf{WikiAnswers}~\cite{WikiAnswers}, questions grouped as
paraphrases;
\textbf{Quora}~\cite{QuoraQP}, question pairs labeled as semantic
duplicates; and
\textbf{ChatGPT-Paraphrases}~\cite{ChatGPTParaphrases}, source texts
paired with LLM-generated paraphrases.
We embed all corpora with all-MiniLM-L6-v2, which produces normalized embeddings.

\subsubsection{Fold selection}
We construct five folds per dataset while preserving the original corpus order. Indexed sets are consecutive blocks of 80,000 embeddings selected from the beginning of the corpus, while query sets are consecutive blocks of 20,000 embeddings selected from its end. Each fold pairs the corresponding indexed and query blocks. Thus, the first fold uses the first 80,000 embeddings for indexing and the last 20,000 for querying; subsequent folds advance inward from both ends. All selected blocks are disjoint. We apply the same selection procedure to every corpus and report the mean and one standard deviation across the five folds.

\subsubsection{Ground truth and thresholds}
We establish ground truth by exact
cosine-similarity search over the indexed embeddings. Write
$\operatorname{sim}_{\cos}(q,X)=\max_{x\in X}\operatorname{sim}_{\cos}(q,x)$
for a query's maximum similarity to the indexed set. A query is positive
when $\operatorname{sim}_{\cos}(q,X)\ge1-\tau$, and negative otherwise. These labels follow the embedding-space membership criterion, independently of the duplicate or paraphrase annotations of the corpora. For each fold and memory budget, we reuse a single stored cover across all evaluated thresholds. This isolates the effect of the query threshold while holding the filter representation fixed.

\subsubsection{Metrics}
We report the false-positive rate (FPR) and false-negative rate
(FNR):
\[
\mathrm{FPR}=\frac{\mathrm{FP}}{\mathrm{FP}+\mathrm{TN}},
\qquad
\mathrm{FNR}=\frac{\mathrm{FN}}{\mathrm{FN}+\mathrm{TP}}.
\]

Let $N>0$ be the total number of queries. With zero false negatives, the
fraction of index accesses saved is $\phi=\mathrm{TN}/N$, since rejected queries
bypass the backend index. When negative queries are present,
\[
 \phi=\frac{\mathrm{TN}}{N}
  =\frac{\mathrm{TN}+\mathrm{FP}}{N}\,(1-\mathrm{FPR}).
\]
Thus, savings equal the fraction of negative queries multiplied by
$1-\mathrm{FPR}$, and the negative-query fraction sets the maximum achievable
saving. We report $100\phi$ as the percentage of accesses saved. If no query is
negative, the savings are zero.

In some evaluations, to distinguish distant negatives from those near the indexed
set, we grouped negative queries by their maximum cosine similarity to an indexed vector: least related ($\operatorname{sim}_{\cos}(q,X)<0.6$), moderately related ($0.6\le\operatorname{sim}_{\cos}(q,X)<0.8$), and most related ($\operatorname{sim}_{\cos}(q,X)\ge0.8$).

\subsubsection{Methods and memory accounting}

We compare different construction methods for selecting the stored caps, 
all of which cover the entire indexed set and allow zero false negatives.
The construction methods heuristically minimize the volume of stored caps,
since we do not estimate the query distribution.

Quantization experiments compare TQ~\cite{TurboQuant, vargaftik2022eden}
and SPQ. Unless otherwise stated,
the filter uses soft anneal with automatic encoding selection
as described in \cref{sec:static-construction}.
% The external baselines and their calibration procedures are described in \cref{sec:eval-comparison}.

Equal-memory comparisons account for each method's query-time
representation, including codes, angles, hash tables, bucket
entries, identifiers, and other required metadata. Memory is
reported relative to the raw FP32 embedding array. The filter
is additional to the backend index, and its actual cap count
depends on the encoding selected for the filter within the byte budget.

For $|X|=80{,}000$ vectors of dimension $d=384$, $M_{\mathrm{raw}}\approx123\ \text{MB}$.

\subsubsection{Implementation and timing}
Our implementation uses Python with a Rust query kernel exposed
through PyO3. We use FAISS~\cite{FAISS} for exact search and the
flat backend index. The implementation is open-sourced and available online~\cite{GitHub}.

End-to-end timing experiments run on one pinned CPU core, one job at a time, and process queries individually. For each memory budget and threshold, we warm both paths and alternate baseline and filtered query passes, reporting the median total query runtime over three repetitions. Filter construction and first-query preparation are recorded separately. Data loading, backend construction, ground-truth computation, correctness checks, and memory accounting are outside the query timers. We measure filter-only query time separately for backend-cost projections. Per-query transformations and backend searches remain part of the timed query paths.

\subsection{End-to-End Filtering Benefit}
\label{sec:system}

We evaluate the benefit of placing a \sysname{} in front of an exact
FAISS flat index. Queries rejected by the filter bypass the index, 
while accepted queries require an exact search. Since the
filter guarantees zero false negatives, the fraction of queries with
no similar indexed vector provides an upper bound on the
reduction in index accesses.

We use soft anneal with automatic encoding selection
(\cref{sec:eval-quantization}), varying the memory budget
($M_{\mathrm{filter}}/M_{\mathrm{raw}}$) and cosine-distance threshold ($\tau$).

\subsubsection{Index accesses saved}
\begin{figure*}[!t]
  \centering
  \includegraphics[width=\linewidth]{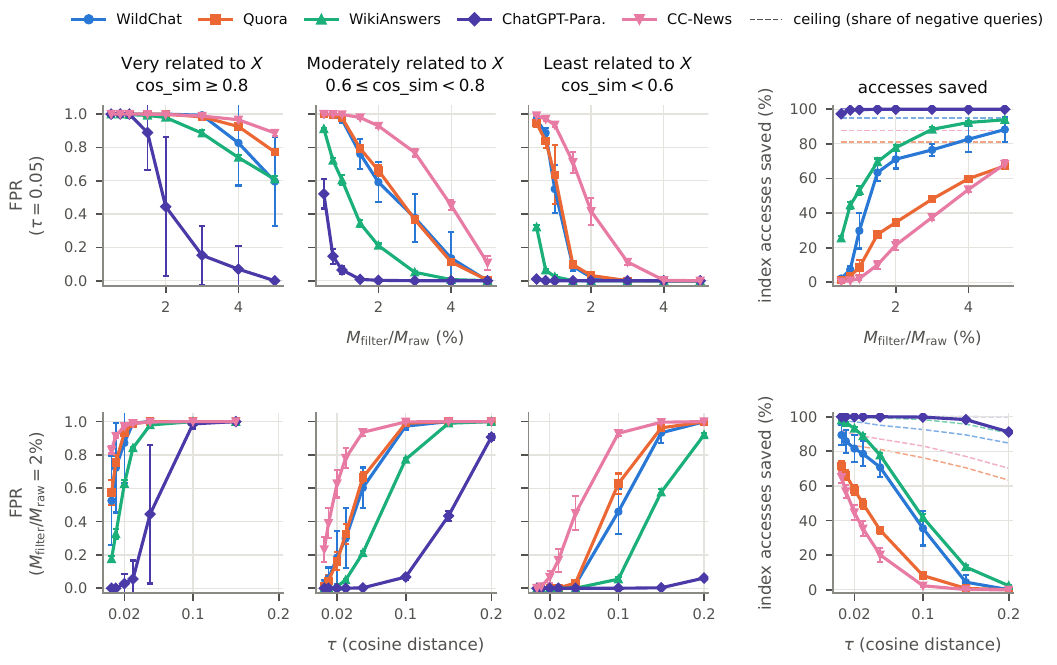}
  \caption{Filtering benefit of \sysname{}.
  \textbf{Top:} memory ratio
  $M_{\mathrm{filter}}/M_{\mathrm{raw}}=0.5$--$5\%$
  at $\tau=0.05$.
  \textbf{Bottom:} cosine-distance threshold
  $\tau=0.005$--$0.2$ at a memory ratio of $2\%$.
  The first three columns report FPR for negative queries grouped
  by their maximum similarity to the indexed set.
  The last column reports index accesses saved; dashed lines show
  each dataset's negative-query fraction.
  Each filter uses automatic encoding selection; a band is not drawn where it has no negative queries.
  Error bars show $\pm1$ standard deviation across five folds.}
  \label{fig:end-to-end}
\end{figure*}

The memory sweep in the top row of \cref{fig:end-to-end} shows that \sysname{} can avoid a substantial share of backend searches while using relatively little memory. Increasing the budget brings savings closer to the ceiling set by each workload's share of negative queries. The proximity breakdown in \cref{fig:end-to-end} helps explain this gap: negatives close to indexed embeddings remain difficult to reject even at the largest evaluated memory budget, while more distant negatives are filtered effectively. This behavior reflects the excess space introduced by geometric summarization.

The bottom row of \cref{fig:end-to-end} examines a similarity threshold sweep while keeping the stored cover and its memory footprint fixed. Increasing $\tau$ expands each cap's acceptance region, increasing the number of queries forwarded to the index.

On most datasets, savings fall below the negative-query ceilings as the threshold becomes more permissive. Queries closest to the indexed embeddings become difficult to reject first, followed by more distant negatives as the acceptance regions expand further.

ChatGPT-Paraphrases retains effective filtering over a wider threshold range because its queries are predominantly farther from the indexed set. The useful operating range thus depends jointly on the requested cosine-distance threshold and the proximity of negative queries to the indexed embeddings.

\subsubsection{Throughput}
To measure the end-to-end impact, we compare an unfiltered FAISS flat index with the same index preceded by \sysname{}. Both configurations use exact range search for queries reaching the backend, while rejected queries bypass that search.

\begin{figure*}[!t]
  \centering
  \includegraphics[width=0.75\linewidth]{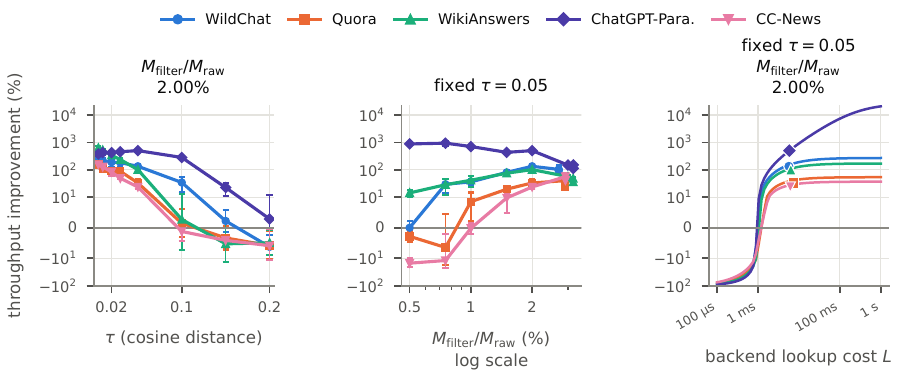}
  \caption{Throughput improvement (\%) by adding a \sysname{} to a flat FAISS index. Left: vs.
  $\tau$ (cosine distance) at $2\%$ of $M_{\mathrm{raw}}$. Middle: vs. the
  memory ratio $M_{\mathrm{filter}}/M_{\mathrm{raw}}$, at $\tau=0.05$.
  Right: modeled improvement versus backend lookup cost, using independently measured filter-only query time and the acceptance rate at $\tau=0.05$ and $2\%$ memory. Markers show measured end-to-end improvement at the measured flat-index cost. Negative improvements denote slowdowns. Error bars in the left and middle panels: $\pm 1$ standard deviation over five folds.}
  \label{fig:throughput}
\end{figure*}

The measured results in \cref{fig:throughput} show that, at $\tau=0.05$, a filter occupying 2\% of the raw embedding memory increases throughput by 116\% on average across the five datasets, ranging from 5\% (CC-News) to 382\% (ChatGPT-Paraphrases), with a median of 56\%. Avoided searches therefore more than compensate for the cost of filtering. Looser thresholds avoid fewer searches and reduce the throughput gain.

The cost projection in the right panel of \cref{fig:throughput} further shows how the value of rejection depends on the backend. More expensive lookups make each avoided search more valuable, allowing the same filter cost and rejection rate to produce a larger throughput gain. The benefit therefore depends on both filtering selectivity and the cost of the operation it bypasses.

\paragraph{Break-even cost}
Let $L_{\mathrm{filter}}$ denote the average filter cost and $L$ a constant backend lookup cost. With rejection rate $\phi$, the average filtered query costs $L_{\mathrm{filter}}+(1-\phi)L$, giving the break-even condition $L>L_{\mathrm{filter}}/\phi$ for $\phi>0$.

The right panel of \cref{fig:throughput} simulates the benefit of applying \sysname{} as a function of the backend cost $L$.

Larger flat indexes provide more search work to avoid, while the same cost model suggests potential benefits for approximate disk-resident or remote indexes and expensive verification stages.

\subsection{Quantization}
\label{sec:eval-quantization}

Direction encoding determines how a fixed memory budget is divided between the number of caps and the precision of their directions. Compact codes permit more caps and potentially tighter covers, while their representation error can increase the angles needed to enclose the indexed vectors. We preserve coverage by recomputing these angles around the decoded directions (\cref{sec:quantized-cover}).

\begin{figure*}[!t]
  \centering
  \includegraphics[width=\linewidth]{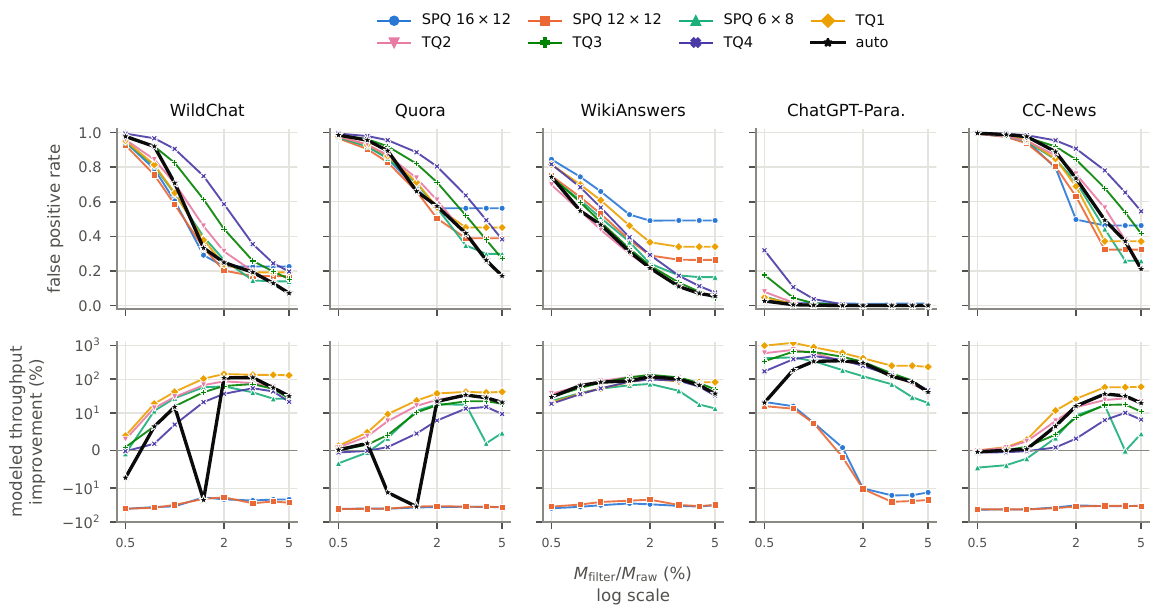}
    \caption{FPR (top) and modeled throughput improvement (\%, bottom) versus $M_{\mathrm{filter}}/M_{\mathrm{raw}}$ at $\tau=0.05$, one column per corpus. Auto (thick black line, stars) and seven fixed encodings are shown: SPQ $16\times12$, $12\times12$, $6\times8$, and TQ1--TQ4, selected by mean rank of accesses saved across corpus--budget cells. The model combines batched filter-query cost and acceptance rates with a mean flat-index lookup cost of $5.29$ ms from end-to-end measurements. The horizontal axis is logarithmic; the bottom vertical axis is symmetric-log, with negative values indicating slowdowns. Curves show means over five folds. Auto selects the feasible encoding with the smallest unexpanded cap volume.}
  \label{fig:quantization}
\end{figure*}

\begin{figure*}[!t]
\centering
\includegraphics[width=\linewidth]{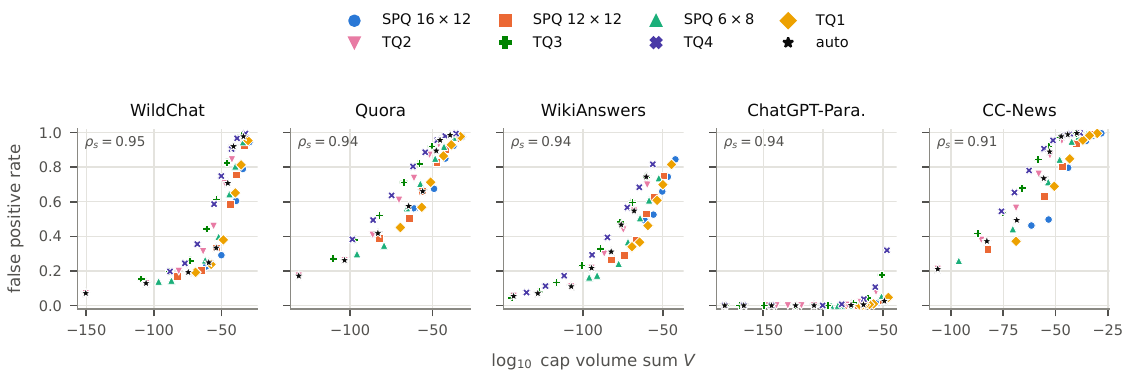}
\caption{FPR versus $\log_{10}V$, the unexpanded cap-volume sum, at $\tau=0.05$, one panel per corpus. Each point averages FPR and $\log_{10}V$ over five folds for one encoding--budget pair, using the seven fixed encodings in \cref{fig:quantization} with the same markers. Auto (stars) repeats the point of the encoding it selects and is excluded from the correlations. Panel annotations give the Spearman correlation over the fixed-encoding points, pooling budgets.}
\label{fig:volume-fpr}
\end{figure*}

\Cref{fig:quantization} compares TQ~\cite{TurboQuant} and SPQ~\cite{jegou2011pq}, alongside automatic encoding selection. SPQ generates its codebooks from a fixed seed, allowing their entries to be reconstructed instead of stored explicitly with the filter.

The memory sweep shows that the best encoding depends on both the workload and the available budget. The automatic configuration remains broadly competitive. Query cost adds another factor to the choice of quantization method. Thus, our default method, auto, may only select from a set of quantizations that were found to be most effective and efficient for this use case (memory aligned TQ and SPQ). The filter chooses the encoding that yields the cover with the smallest unexpanded cap-volume objective $V$ (\cref{sec:static-construction}). We empirically found that a smaller $V$ correlates with a lower FPR in our evaluations. Among the plotted encodings, TQ1 gives the highest modeled throughput improvement in 34 of the 40 corpus--budget cells; TQ2, TQ3, or auto lead in the others. The model uses batched filter-query times, whereas \cref{fig:throughput} times queries individually, so the modeled gains are not directly comparable to the measured ones.

At $0.5\%$ memory, FPR is near one on Quora and CC-News, whereas ChatGPT-Paraphrases already has auto FPR $0.026$, falling to $0.002$ at $1\%$. At $2\%$, SPQ $16\times12$ and $12\times12$ have lower FPR than TQ4 on WildChat, Quora, and CC-News, but both incur modeled slowdowns on all five corpora. The auto configuration selects the slower SPQ methods when they beat TQ on volume. Thus, when using auto, the quantization options should be carefully selected to balance FPR and speed.

\Cref{fig:volume-fpr} shows a positive association between unexpanded volume and FPR on every corpus ($\rho_s=0.91$--$0.95$). Budget drives both quantities: correlating the seven fixed encodings within each budget and averaging over the eight budgets gives only $-0.03$ (CC-News) to $0.87$ (WikiAnswers), and 14 of the 40 corpus--budget correlations are negative. Encodings also differ in FPR at comparable volumes, so the pooled association supports $V$ as a heuristic, not a reliable ordering of encodings at fixed memory.

\subsection{Construction Methods}
\label{sec:eval-construction}

\begin{figure*}[!t]
  \centering
  \includegraphics[width=\linewidth]{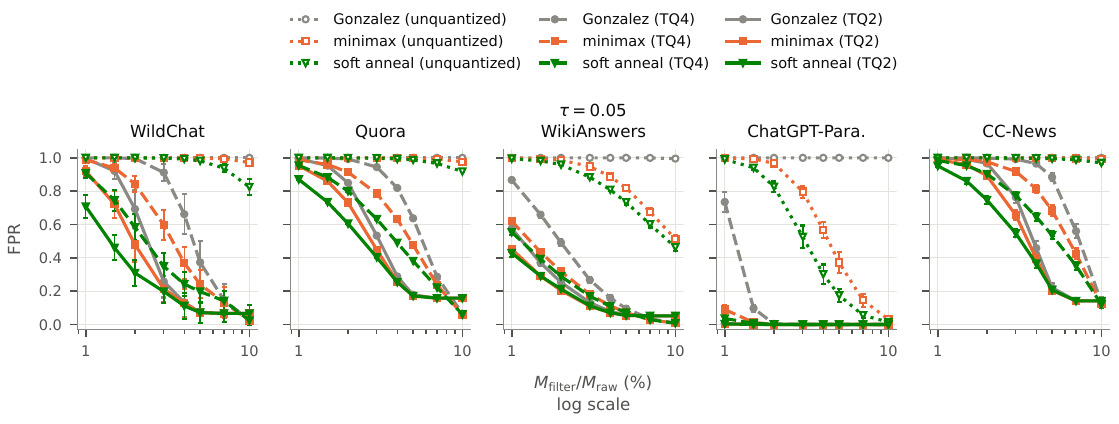}
  \caption{FPR versus memory ratio $M_{\mathrm{filter}}/M_{\mathrm{raw}}$ at cosine-distance threshold $\tau=0.05$, for Gonzalez, minimax, and soft anneal. Each method is shown with unquantized FP32, TQ4, and TQ2 directions. The memory axis is logarithmic. Error bars: $\pm 1$ standard deviation over five folds.}
  \label{fig:construction}
\end{figure*}

\Cref{fig:construction} compares Gonzalez, minimax, and soft anneal to examine how the construction
method affects the FPR of \sysname{} as a function of the memory ratio. Minimax may only decrease the maximum angle Gonzalez provides, while soft anneal may only reduce the unexpanded cap-volume objective $V$ minimax provides. The figure compares unquantized FP32 directions with TQ2 and TQ4 directions. While TQ2 achieves a lower FPR by storing more caps, once it stores a cap per indexed vector (clear in the Quora workload), it stops improving. Eventually, TQ4 achieves a lower FPR due to a lower quantization error.

Construction timings are measured sequentially on one pinned CPU core and include all stages used by each method. Minimax includes its
Gonzalez initialization, while soft anneal additionally includes minimax. For 80,000 vectors without applying quantization, total construction time $\pm$ standard deviation averaged ($5.88 \pm 3.48$) seconds for Gonzalez, ($15.02 \pm 7.51$) seconds for minimax, and ($355.58 \pm 128.82$) seconds for soft anneal, making minimax a reasonable middle ground between runtime and quality.

\subsection{Comparison Against Baselines}
\label{sec:eval-comparison}

We compare methods at equal query-time memory budgets to determine where geometric summaries provide useful filtering relative to alternative representations. Curves begin at the first sampled feasible budget.

\begin{figure*}[!t]
\centering
\includegraphics[width=\linewidth]{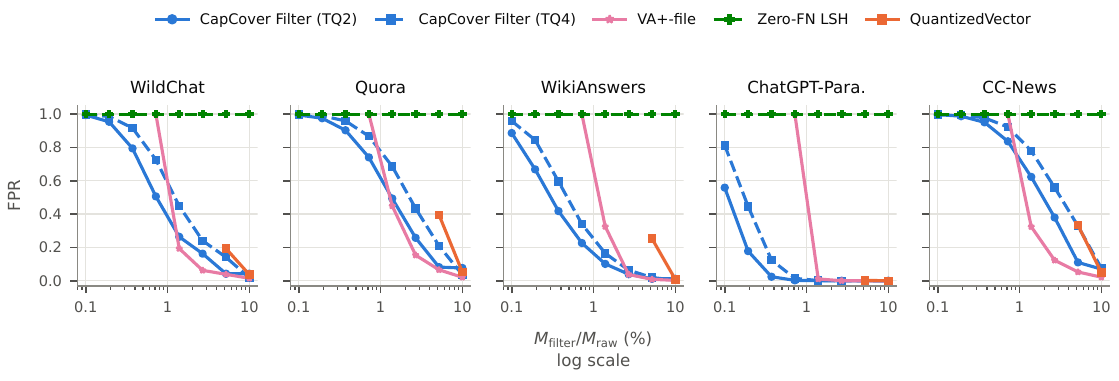}
\caption{Mean FPR of zero-false-negative methods, at cosine-distance threshold $\tau=0.05$. Curves begin at the first sampled feasible budget.}
\label{fig:equal-memory}
\end{figure*}

We compare against VA+-file~\cite{ferhatosmanoglu2000vaplus}, which uses lower bounds to KLT-rotated quantization cells; zero-FN LSH schemes~\cite{sankowski2017nofn,pacuk2016nofn}, which preserve close pairs through adjacent floor-hash cells; and a quantized point set using RaBitQ~\cite{gao2024rabitq} codes with conservative bounds on representation error.

\Cref{fig:equal-memory} shows that \sysname{} provides useful rejection at compressed operating points where the other evaluated one-sided representations remain ineffective or do not fit. Both zero-FN LSH schemes fit within small budgets but have FPR~1: their representations preserve close pairs, yet provide no useful exclusion on these workloads. The advantage of the geometric cover is consistent with its ability to summarize multiple indexed vectors within each stored region, sharing the representation cost across their members.

As budget increases, point-oriented representations gain enough precision to reject queries effectively. VA+-file can then achieve lower FPR than \sysname{}, while the quantized point set becomes feasible at larger budgets. The resulting crossover identifies the benefit of geometric summaries for rejection when memory is too restricted to represent individual vectors effectively.

\paragraph{Methods that allow false negatives}

\begin{figure}[h!]
  \centering
  \includegraphics[width=\columnwidth]{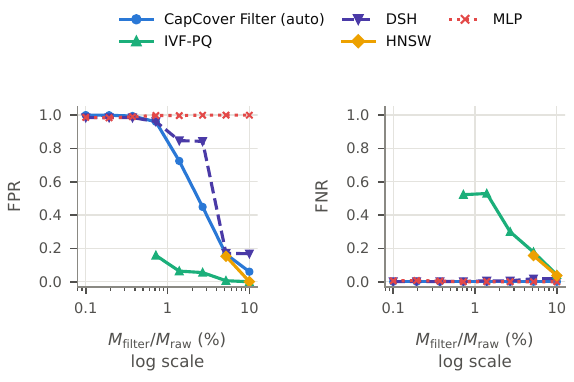}
  \caption{FPR (left) and FNR (right) versus $M_{\mathrm{filter}}/M_{\mathrm{raw}}$ on Quora at $\tau=0.05$, with \sysname{} (zero FNR) as reference. Curves begin at the first sampled feasible budget.}
  \label{fig:equal-memory-nonzero-fn}
\end{figure}

In \cref{fig:equal-memory-nonzero-fn} we compare four baselines that may report a true member as absent. DSH~\cite{aumuller2018distance} stores distance-sensitive hash signatures and accepts a query whose signature collides with a stored one; unlike \sysname{}, it fixes $\tau$ at construction. An \emph{MLP} regresses the distance to the nearest indexed vector and accepts when the prediction is at most $\tau$. IVF-PQ~\cite{jegou2011pq} and HNSW~\cite{malkov2020hnsw} accept when their approximate search returns a vector within $\tau$.

On Quora, IVF-PQ attains low FPR from 0.7\% memory but misses about half of the true members there (FNR 0.52), and HNSW fits only from about 5\%, where its FNR is 0.16. DSH keeps FNR below 0.02, but its FPR stays above CapCover Filter's from 1.4\% memory on and levels off at about 0.17, whereas CapCover Filter reaches 0.06 at 10\% with no false negatives. MLP accepts nearly every query at every budget and therefore provides no filtering.

\begin{comment}    
\subsection{Limitations}
\label{sec:limitations}
Our evaluation covers the static setting. Insertions are supported, but their effect on FPR and cost is not evaluated here. All corpora are embedded with a single model (all-MiniLM-L6-v2, $d=384$) under cosine distance, with 80,000 indexed vectors per fold. We do not study how cover quality changes with dimension, embedding model, or substantially larger indexed sets.
A query scans the entire filter, so its cost grows linearly with the memory budget. The filter improves throughput only when the backend lookup is expensive enough to satisfy the break-even condition $L>L_{\mathrm{filter}}/\phi$ derived above. As budgets grow, point-oriented representations such as VA+-file can achieve lower FPR (\cref{sec:eval-comparison}). Finally, soft anneal takes minutes rather than seconds to construct (\cref{sec:eval-construction}), so its benefit depends on serving enough queries between rebuilds.
\end{comment}

\section{Conclusion}
\label{sec:conclusion}
\label{sec:when}

We introduced \sysname{}, a compact cap cover that rejects semantic nonmatches before index search while guaranteeing zero false negatives. Construction and encoding choices affect selectivity and cost, while conservative angle remeasurement preserves coverage.

The results distinguish geometric quality from system performance. More accurate encodings can lower FPR yet reduce throughput, as seen with SPQ $16\times12$ and $12\times12$ at $2\%$ memory. Encoding selection should therefore account for backend cost: the best representation depends on whether avoided searches compensate for its scan cost. Unexpanded volume is a useful objective, but its strong association with FPR across budgets weakens when comparing encodings at fixed memory. 

Representative queries could help select among candidate encodings, with coverage retaining the independent correctness guarantee. Smaller volume alone does not guarantee better rejection on a particular workload. Geometric summarization is most valuable where multiple vectors can share a stored region; point-oriented representations become competitive as budgets grow. Construction introduces a further tradeoff: additional refinement must be amortized over subsequent queries, making the query volume between rebuilds relevant alongside filter size and scan cost. Future work includes encoding selection informed by backend cost, construction tailored to expected query volume, and space bounds for clustered embeddings.

\begin{comment}    
\section{AI-Generated Content Acknowledgement}
OpenAI's ChatGPT (mainly GPT-6 Astra) was used to review wording and to assist with revisions. Anthropic's Claude Code (mainly Sonnet 5.5) was used to optimize the implementation (especially the Rust kernels), as well as to implement experiments specified by the authors, supporting the implementation and evaluation in Sections III and V. In addition, we utilized it to explore the raw data to find additional insights and patterns.
The authors take full responsibility for the correctness and originality of the content.
\end{comment}

\bibliographystyle{IEEEtran}
\bibliography{citations}

\end{document}